\documentclass[preprint]{aastex}
\usepackage{emulateapj5,graphicx}
 
\slugcomment{AJ accepted}

\shorttitle{Initial Speed of Knots in the Plasma Tail}

\shortauthors{Yagi et al.}
\begin{document}

\title{Initial Speed of Knots in the Plasma Tail of C/2013 R1(Lovejoy)}

\author{Masafumi Yagi\altaffilmark{1,2},
Jin Koda\altaffilmark{3},
Reiko Furusho\altaffilmark{2,4},
Tsuyoshi Terai\altaffilmark{2,5},
Hideaki Fujiwara\altaffilmark{5},
Jun-Ichi Watanabe\altaffilmark{2}}

\altaffiltext{1}{email:YAGI.Masafumi@nao.ac.jp}
\altaffiltext{2}{
National Astronomical Observatory of Japan,
2-21-1, Osawa, Mitaka, Tokyo, 181-8588, Japan}
\altaffiltext{3}{Department of Physics and Astronomy, 
Stony Brook University, Stony Brook, NY 11794-3800, USA 
}
\altaffiltext{4}{Tsuru University, 3-8-1, Tahara, 
Tsuru, Yamanashi, 402-0054, Japan}
\altaffiltext{5}{Subaru Telescope, 650 North A'ohoku Place, Hilo,
Hawaii 96720, USA}

\begin{abstract}
We report short-time variations in the plasma tail of C/2013 R1(Lovejoy). 
A series of short (two to three minutes) exposure images 
with the 8.2-m Subaru telescope shows faint details of filaments and
their motions over 24 minutes observing duration.
We identified rapid movements of two knots in the
plasma tail near the nucleus ($\sim$ 3$\times 10^5$ km).
Their speeds are 20 and 25 km s$^{-1}$ along the tail
and 3.8 and 2.2 km s$^{-1}$ across it, respectively. 
These measurements set a constraint on an acceleration model 
of plasma tail and knots as they set the initial speed
just after their formation.
We also found a rapid narrowing of the tail. 
After correcting the motion along the tail, 
the narrowing speed is estimated to be $\sim$ 8 km s$^{-1}$.
These rapid motions suggest the need for high 
time-resolution studies of comet plasma tails with 
a large telescope.
\end{abstract}


\section{Introduction}

Plasma tails of comets and their time variations potentially provide
crucial information on solar winds and magnetic fields in the solar system
\citep[e.g.,][]{Niedner1982,Mendis2006,Downs2013}.
Short-time variations in plasma tails, however, are not yet fully
understood. 
Indeed, most previous studies observed tails and structures 
at far distances ($>10^6$ km from the nucleus) 
with a time resolution of an order of an hour.

Regarding the speed of movement along the tail, 
\citet{Niedner1981} studied 72 disconnection events (DEs)
of various comet tails and found $\lesssim$ 100 km s$^{-1}$
at $\lesssim 10^7$ km from the nuclei.
Their initial speeds before DEs are around 44 km s$^{-1}$ and
the typical acceleration is 21 cm s$^{-2}$.
\citet{Saito1987} analyzed a knot in the plasma tail of 
comet 1P/Halley and derived its average velocity of 58 km s$^{-1}$
at 4--9 $\times 10^5$ km from the nucleus.
\citet{Kinoshita1996} observed C/1996 B2(Hyakutake)
and measured the speed of a knot of 99.2 km s$^{-1}$ 
at 5.0 $\times 10^6$ km from the nucleus.
\citet{Brandt2002} investigated DE of
C/1995 O1 (Hale-Bopp) and obtained the speed of $\sim$ 500 km s$^{-1}$
at $\sim$ 7$\times 10^7$ km from the nucleus.
\citet{Buffington2008} analyzed several knots in 
comets C/2001 Q4(NEAT) and C/2002 T7(LINEAR) 
using the Solar Mass Ejection Imager.
They found the speed to be 50--100 km s$^{-1}$
around $10^6$ km from the nucleus.

These previous studies 
did not catch the moment immediately after the formation of knots or 
the detachment of knots from the tail.
The initial speed at these critical times was only an extrapolation from 
later observations relatively far away.
In this letter, we report detections of knots in the plasma tail
3$\times 10^5$ km away from the nucleus of C/2013 R1(Lovejoy)
and a direct measurement of their initial motions.
We adopt the AB magnitude system throughout the paper.

\section{Data}

The comet was observed on 2013 December 4 (UT) using
the Subaru Prime Focus Camera \citep[Suprime-Cam;][]{Miyazaki2002}
mounted on the Subaru Telescope at Mauna Kea (observatory code 568%
\footnote{\url{http://www.minorplanetcenter.net/iau/lists/ObsCodesF.html}}).
The camera consists of a 5 $\times$ 2 array of 2k $\times$ 4k CCDs.
The pixel scale is 0.202 arcsec pixel$^{-1}$.
The field of view is about 35$\times$28 arcmin.
We used two broadband filters:
W-C-IC (I-band: center=7970 \AA, full width at half maximum (FWHM)=1400\AA) and 
W-J-V (V-band: center=5470 \AA, FWHM=970\AA) filters.
Both bands trace the plasma tail.
I-band includes predominantly H$_2$O$^+$ line emissions,
while V-band includes CO$^+$ and H$_2$O$^+$ line emissions.

The observation log is given in Table \ref{tab:obslog}.
The total observing time of 24 minutes was spent 
after main science targets of the observing run were set.
The start time of the exposures  have an 
uncertainty of about 1 second.
The position angle and the pointing offset were adopted
to catch the comet nucleus at the bottom-left corner 
and to have the tail run diagonally across the
field-of-view so that the maximum extent of the tail is 
framed in each exposure.

The Subaru telescope's non-sidereal tracking mode\citep{Iye2004}
was used so that the comet was always observed at the same position 
on the CCD array.
For the observing run, the comet's coordinates were calculated 
using the NASA/JPL HORIZONS system%
\footnote{\url{http://ssd.jpl.nasa.gov/horizons.cgi}}
with the orbit element of JPL\#22. 
We used an ephemeris of one-minute step.
Unfortunately, the values of JPL\#22 were not recorded. 
In the following, we instead used newer orbital elements
(JPL\#55). The values are given in Table \ref{tab:orbit}.
The positions calculated from JPL\#22 and JPL\#55
have an offset by 0.34 arcsec in right ascension and -1.88 arcsec 
in declination but show no drift during the observation.
The offset of the absolute celestial coordinate 
does not affect in this study, since our analysis is on relative 
position of structures inside of the comet.
At the time of the observations, the 
observercentric and heliocentric distance 
to the comet were 
0.5523--0.5526 and 0.8812--0.8811 au,
respectively. 
In the sky projection, the conversion from the angular to 
the physical scales was 400.6--400.8 km arcsec$^{-1}$. 
Considering the phase angle (sun-target-observer angle) of 83.5 degrees
at the time of the observations, 
we adopt the physical scale along the tail of 
403.3 km arcsec$^{-1}$ in the following discussion.
The heliocentric ecliptic coordinate of the comet nucleus was 
($\lambda$,$\beta$)=(87.5,30.7).
The comet was located before the perihelion passage, 
and its heliocentric velocity was -12.6 km s$^{-1}$.

The seeing size is estimated from short (two-second) exposures
and was 1.0 and 1.1 arcsec in I and V-bands, respectively.
The movement of the comet in the celestial coordinate 
was about (dRA$\times$cos(D)/dt,d(D)/dt)=(283 -- 284,-115) arcsec hr$^{-1}$
during the observations.
Due to the non-sidereal tracking, stars move in an exposure, but
this motion does not affect the measurement of the seeing size
significantly since the shift is only about 0.17 arcsec in 
a two-second exposure.

\begin{figure}
\includegraphics[scale=0.45,bb= 0 0 1042 837]{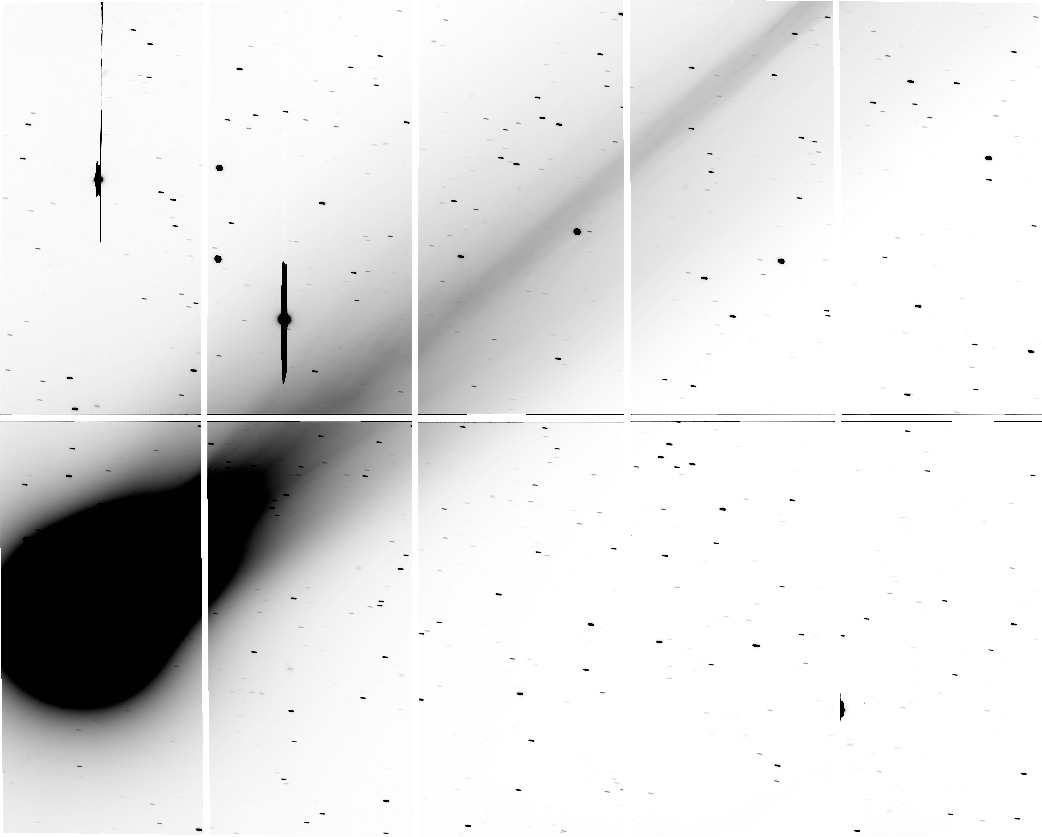}
\caption{An example of the data (V1; see Table \ref{tab:obslog}).
The bright part is shown as black.
Suprme-Cam consists of 10 CCDs, and the gaps between chips are 
shown in white.
The field of view is 34$\times$27 arcmin.
}
\label{fig:V1}
\end{figure}

The data was reduced in a standard manner; the steps include
overscan subtraction, crosstalk correction \citep{Yagi2012},
flat fielding using twilight flat, and distortion correction.
The relative flux and relative position among the CCDs 
are calibrated using other dithered datasets
taken in the same night.
The mosaicked image of V1 (Table \ref{tab:obslog}) 
is shown as Figure \ref{fig:V1} as an example.

The flux was calibrated against stars in the field 
using the Eighth Data Release of the Sloan Digital Sky Survey 
\citep[SDSS DR8;][]{DR8} catalog
in the same way as in \citet{Yagi2013}. 
We first used SExtractor \citep{Bertin1996} for object detection in the
Suprime-Cam images and astrometric calibration was performed 
to the center of the elongated stars against the Guide Star Catalog 2.3.3
\citep{Lasker2008}.
Then, we measured aperture fluxes at each position of 
SDSS star whose r-band magnitude is 16$<$r$<$20.
The aperture radius was 40, 50, and 60 pixels,
and we used 
\begin{equation}
F=F(40)-\left[ (F(60)-F(50)) \times \frac{40^2}{60^2-50^2}  \right]
\end{equation}
as the background-corrected aperture flux of an elongated star,
where $F(r)$ is the aperture flux of $r$ radius.
Photometric zero point was estimated from the instrument flux $F$ and 
the catalog magnitude converted to the Suprime-Cam AB system.
The color conversion coefficients 
from SDSS to Suprime-Cam magnitudes are given in Table \ref{tab:colorcoeff}.
The K-correct \citep{Blanton2007} v4 offset 
for SDSS\footnote{\url{http://howdy.physics.nyu.edu/index.php/Kcorrect}} 
is applied; $m_{AB}-m_{SDSS}$=0.012, 0.010 and 0.028 for $g$, $r$ and $i$, 
respectively.
The number of stars used for calibration was
$\sim 290$ in V and $\sim 410$ in I.
The measured photometric zero-point of chip 2 (reference chip)
was 27.36(I) and 27.15(V) AB mag per 1 ADU s$^{-1}$,
and the peak-to-peak variation among exposures were smaller than 0.02 mag
in each band.
These zero points are comparable to the ones derived typically under
photometric conditions, and thus we regard the data as
having been obtained under photometric conditions.

\section{Result}

\subsection{Image Processing}
To investigate the fine structures in the plasma tail, 
additional image processing was applied.
First, we rotate the image counterclockwise by 50 degrees 
so that the plasma tail aligns to the y-axis,
and the region around the plasma tail within full width of 
1000 pixels (3.4 arcmin) was extracted.
The position angle of the image to the north is the same 
among the nine images (160.0 degree).
In the coordinate, the sun lay 
in the direction of (x,y)=(-0.014,+1.000).
The nuclei position was measured in short exposures (I2 and V2).
The relative positional consistency among the images
depends on the accuracy of the non-sidereal tracking mode of the
Subaru telescope. As we used the ephemeris of a one-minute step,
enough accuracy is guaranteed.

We then applied an unsharp masking technique to each image;
we convolved the extracted images with various sizes of Gaussian filter 
($\sigma$=10, 20, 30, 50, 75 pixels) 
and subtracted the smoothed image from the original ones.
The details are presented in Appendix A.
This processing removes larger structures 
and enhances fine structures.
To differentiate the processing with different kernel sizes, 
we refer to each
as highpass${\it \sigma}$, e.g., highpass10 when processed with
the Gaussian with $\sigma$=10 pixels.
Bright stars in the background contaminate small features 
associated physically with the cometary tail (e.g., knots), 
and hence, we iteratively masked them.
Finally, we binned the image by 5$\times$5 pixels (1.01 arcsec square).
Figure \ref{fig:knots} shows six longer-exposure images 
after the processing.
These images show the gaps between CCD chips as gray bands.
Brightening/darkening around chip edges are artifacts.
These do not affect the following analyses.
 
\subsection{Moving Knots}
\label{sec:knots}

\begin{figure}
\includegraphics[scale=0.5,bb= 0 0 783 1112]{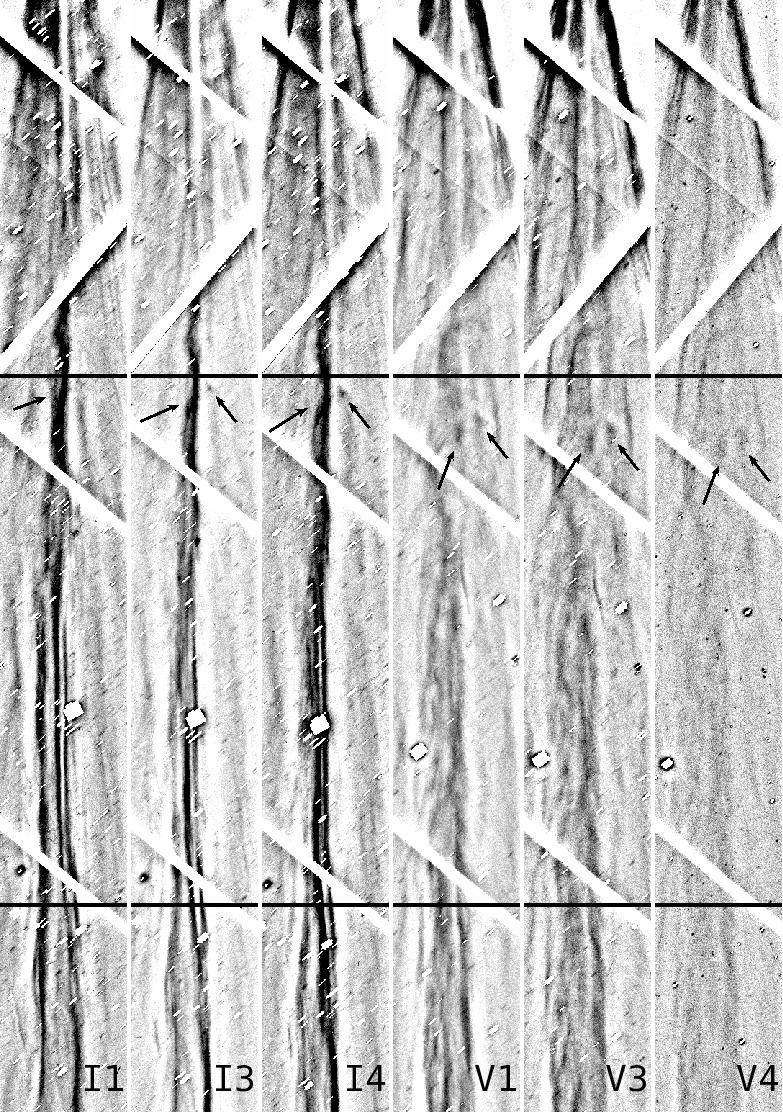}
\caption{
Unsharp masked (highpass50)
and 5$\times$5 pixel binned images.
The gaps between CCDs are seen in gray.
Long exposures, I1, I3, I4, V1, V3 and V4 are displayed
from the left to the right.
The sun and the nucleus are toward the top.
The distance from the nucleus is shown as horizontal lines
at 3$\times$10$^5$ km and 6$\times$10$^5$ km.
Two knots are indicated by arrows.}
\label{fig:knots}
\end{figure}

We can visually see two knots moved downstream
around the distance of $\sim$ 3$\times 10^5$ km from the nucleus 
(indicated by arrows in Figure \ref{fig:knots}).
The closeup images are shown as Figure \ref{fig:closeup}.
One of the knots first appears to be connected to the global filamentary 
structures running along the tail.
It later becomes detached from the structures.
The positions 
of the knots are measured
by running SExtractor \citep{Bertin1996} on 
the 5$\times$5 binned 
unsharp masked images.
We measured in highpass30, highpass50, highpass70, 
and highpass90 images.
As examined in Appendix B, an unsharp masking with 
a smaller kernel may introduce larger
positional errors. Meanwhile, the knot may not be detected 
after an unsharp masking with a large kernel 
because it is buried in global features.
We therefore adopted the position from the largest kernel 
in which the peak of the knot is detected.
The root mean squares (rms) of the position was $\sim 6\times 10^2$ km.

Distances from the nucleus (along the tail; y-axis) 
and offsets from the tail axis (perpendicular to the tail; x-axis) 
are plotted as a function of time in Figure \ref{fig:speed}.
20 and 25 km s$^{-1}$ along the tail
and 3.8 and 2.2 km s$^{-1}$ across the tail, respectively.
The motion should make the shape of the knots elongated 
in a two-minute exposure
by 6--7 arcsec along the tail and 1.3 or 0.7 arcsec across the tail.
Since the size of the knots is at least 10--15 arcsec in diameter,
elongation by the motion 
had little effect on the size estimation.

These knot motions show tilts from the central axis of 
the tail by about 0.1--0.2 radians (6--11 degrees). 
We note that the direction to the Sun is 0.014 radians (0.8 degrees) 
from the axis, and therefore, the knot motions are not in 
a perfect alignment to the direction to the Sun.

The distances to the knots from the nucleus change almost linearly
with time in Figure \ref{fig:speed}. The effect of acceleration is
therefore small in the short duration of our observations 
(23 minutes). The clear detection of the spatial motions, 
however, indicates the potential for direct measurements 
of the acceleration in future.
If we can take the accelerations of previous measurements of other
comets by \citet{Niedner1981} (21 cm s$^{-2}$), 
the expected change in the speed is only 0.3 km s$^{-1}$ in
23 minutes.
If we adopt 17 cm s$^{-2}$, which is calculated from
Figure 4 of \citet{Saito1987} as 2.4 km s$^{-1}$ increment 
is seen in four hours, the expected change in the speed during
our observation is even smaller (0.2 km s$^{-1}$). 
Therefore, a sequence of few hours
observations of a knot would 
permit us to measure the acceleration.

\subsection{Change of tail width in I-band}

We found that the width of the plasma tail also rapidly changed.
In the analysis, the original images before 
the unsharp masking were used.
In Figure \ref{fig:Itail} and 
Figure \ref{fig:width} show spatial surface brightness 
profiles across the tail and their time variation.
Since we have measured the motions of the knots 
(presumably the motion of the tail),
we can track the variation in the tail width at a comoving position
as the tail would flow at the speed of $\sim$ 22 km s$^{-1}$.
In a sequence of images (I1, I3, and I4), 
we measured a profile at the distance of 6$\times$10$^5$ km in I1. 
We then tracked the location of the material initially at this distance, 
but at later times using the flow speed, and plotted the profiles of 
(presumably) the same material in I3 and I4.
The tail width is determined at the surface brightness of 
21.3 AB magnitude arcsec$^{-2}$ in I-band.
It narrowed from 2.4$\times 10^4$ to 2.0$\times 10^4$, and 
then to 1.8$\times 10^4$ km for I1, I3, and I4, respectively.

From this measurement, the speed of the narrowing motion across 
the tail is $\sim$ 8 km s$^{-1}$ at each edge. 
The center of the light distribution, meanwhile,
did not show any notable change (Figure \ref{fig:width}). 
This narrowing speed is larger than those of the knots in 
the same direction perpendicular to the tail 
(3.8 and 2.2 km s$^{-1}$; Section \ref{sec:knots}).
This apparent difference could be attributed to a projection of 
the motion of the knots in the sky.

If we assume every fine structure (filamentary structure) 
in the tail is moving at $\sim$ 8 km s$^{-1}$
perpendicular to the tail, the amount of spatial shift 
would be about 2.5 arcsec within the 120-second exposures. 
On the other hand, the fine structures appear to be a typical 
5 --10 arc sec width in short exposures (I2, V2). Therefore, 
the blur due to the motions during a single 120-second exposure
should not affect the positional measurements of the structures much.

\section{Discussion and Summary}

The initial speeds of two moving knots ($\sim$ 22 km s$^{-1}$) are 
significantly slower
than the ones measured by \citet{Niedner1981}
(44 km s$^{-1}$; rms of 10.9 km s$^{-1}$).
This speed is also smaller than that measured in comet 
Halley by \citet{Saito1987} (58 km s$^{-1}$),
who suggested that the velocity was
constant at 4--9 $\times 10^5$ km from the nucleus.

Though it is not clear what the dominant factor 
of the initial speed of knots is, 
we can compare several parameters of the comets.
As the data used by \citet{Niedner1981} include
information from various comets, we compare parameters with 
the single case of the comet Halley by \citet{Saito1987}.
At the observation by \citet{Saito1987},
the heliocentric distance to comet Halley was 1.016 au, 
the heliocentric ecliptic coordinate was 
($\lambda$,$\beta$)=(30.1,8.8),
and the heliocentric velocity was -26.5 km s$^{-1}$.
Compared with  C/2013 R1(Lovejoy) in this study,
a part of the difference in the initial speed 
may be explained by the difference in the 
heliocentric velocity of the nucleus,  
-12.6 km s$^{-1}$ versus -26.5 km s$^{-1}$,
if we assume that the initial speed of the knots 
might be comparable in heliocentric frame.
It, however, does not fully explain $\sim$ 40 km s$^{-1}$ difference.
Another difference is heliocentric ecliptic latitude,
30.7 vs 8.8, which may result in the difference in the speed of 
the solar wind at the comet position.
Yet another point is that
we have compared the speeds of our relatively faint and small knots 
with more prominent knots/kinks and DEs in the previous studies. 

The relevance of this comparison might be debated in light of future 
studies. In addition, we analyzed only one comet tail observed in a relatively
short duration. A more systematic investigation is obviously needed as
to the distribution of the initial speed of the tail as a function of
the heliocentric velocity, the heliocentric distance,
and the ecliptic position of the comet.

In summary, we found two knots that were just formed at 3$\times 10^5$ km
from the nucleus of  C/2013 R1(Lovejoy).
Their initial speed was smaller than the ones measured in previous studies,
and a physical interpretation requires a more statistically significant 
sample at various heliocentric positions.
We also found a rapid variation in the tail width in seven minutes,
which implies a rapid change in ambient solar winds and magnetic field.
These results strongly suggest that the variations in comet plasma tails,
especially in their fine structures, require high time resolution 
observations with a large aperture telescope such as Subaru.

\acknowledgments
This work is based on observations obtained with the Subaru Telescope. 
We acknowledge David Thilker, Alexandre Y. K. Bouquin, 
Fumiaki Nakata, and Yutaka Komiyama for their help.
We thank Miriam Forman for suggestive comments.
We also thank anonymous referee for useful comments.
This work has made use of 
SDSS DR8 database,
HORIZONS ephemerides service in NASA/JPL,
and
the computer systems at Astronomical Data Analysis Center of NAOJ.
JK is supported by the NSF through grant AST-1211680 and by NASA 
through grants NNX09AF40G, NNX14AF74G,
a Herschel Space Observatory grant, 
and a Hubble Space Telescope grant.

\onecolumn

\begin{figure}
\includegraphics[scale=0.45,bb= 0 0 1026 556]{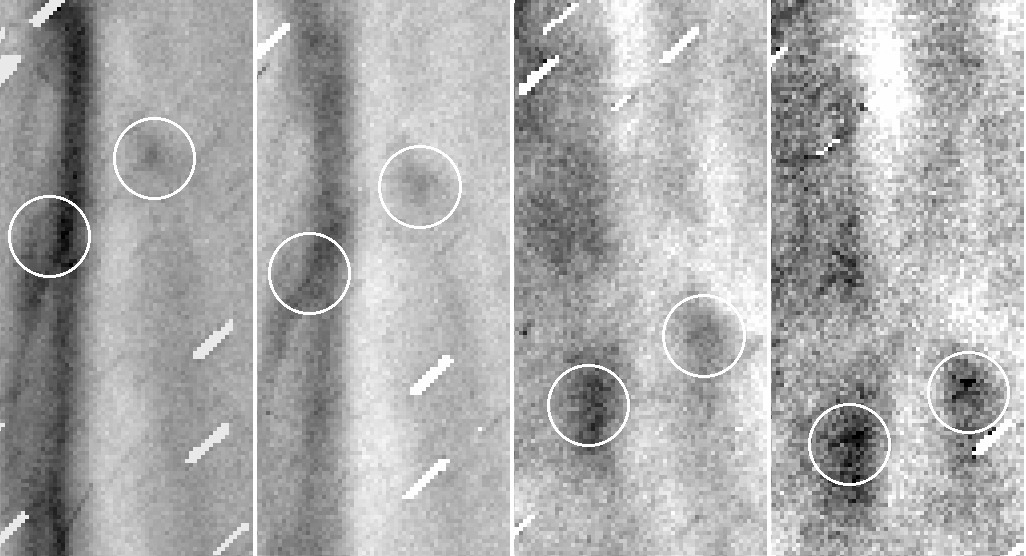}
\caption{
Closeup images around knots.
I3, I4, V1, and V3 are displayed from the left to the right.
I3 and I4 are highpass50, and V1 and V3 are highpass70 filtered.
The positions used in the Figure \label{fig:speed} are indicated
with white circles.}
\label{fig:closeup}
\end{figure}

\begin{figure}
\includegraphics[scale=0.6,bb= 0 0 572 762]{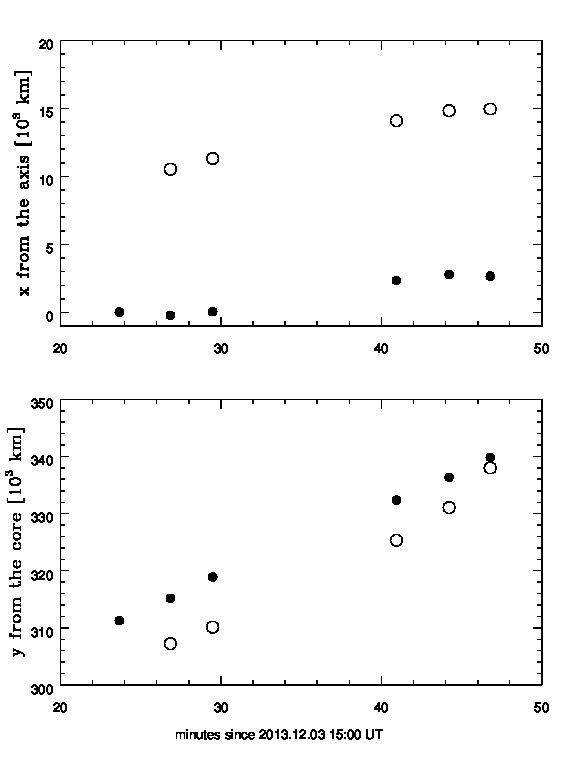}
\caption{The position of moving knots
in Figure \ref{fig:knots} and \label{fig:closeup} as a function of time.
Open and filled symbols correspond to the position of each knot.
The top panel shows the tangential distance from the nucleus 
measured along the tail, and the bottom panel shows the distance from
the axis. 
The abscissa shows the time in minutes since 2013 Dec. 4 at 15:00 UT.
}
\label{fig:speed}
\end{figure}

\begin{figure}
\includegraphics[scale=0.6,bb= 0 0 795 796]{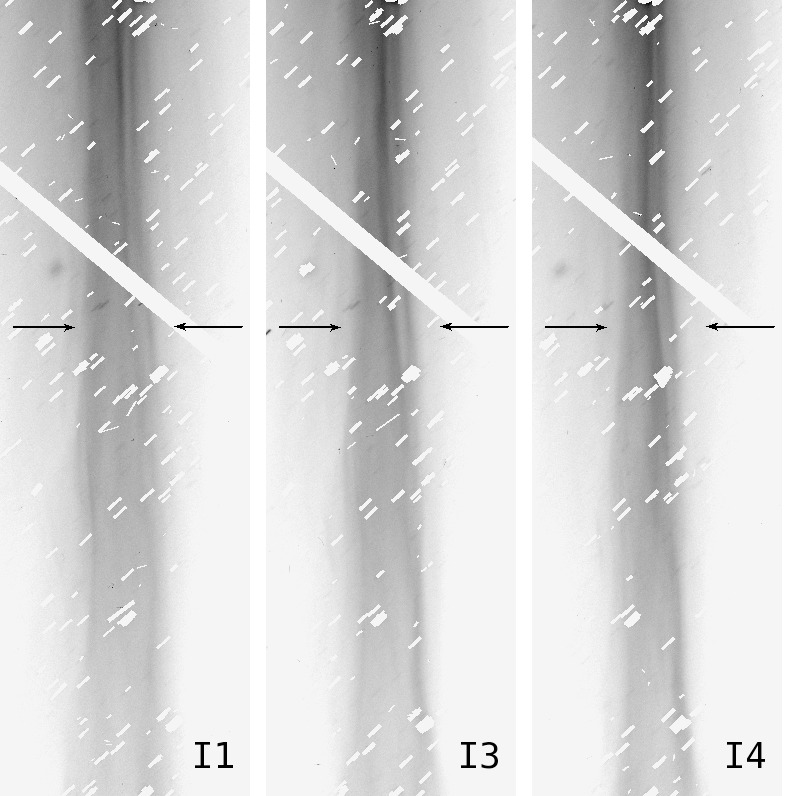}
\caption{The change in the width of the tail in I-band frames.
The images before unsharp masking are shown.
The y-positions are shifted assuming a flow of 22 km s$^{-1}$.
The black arrows show 6$\times$10$^5$ km from the nucleus in I1,
where the profile is measured (Figure \ref{fig:width}).
}
\label{fig:Itail}
\end{figure}

\begin{figure}
\includegraphics[scale=0.6,bb= 0 0 768 574]{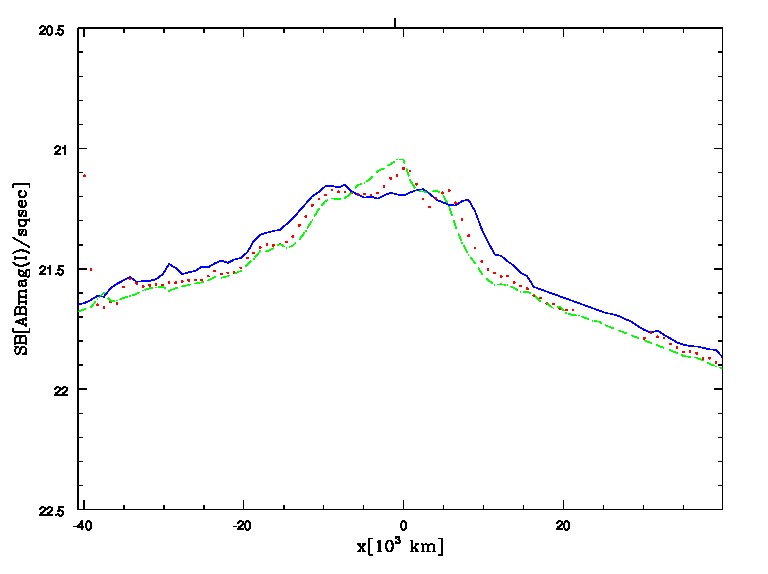}
\caption{The change in the spatial surface brightness profile 
across the tail in I-band frames.
Blue solid, red dotted and green broken lines corresponds to 
I1, I3, and I4, respectively.
The y-position was 6$\times$10$^5$ km from the nucleus in I1,
and shifted assuming 22 km s$^{-1}$ movement along the tail.
}
\label{fig:width}
\end{figure}

\begin{table}
\begin{tabular}{ccccc}
\hline
tag & filter & UT(start) & exptime(sec) \\
\hline
I1 &I & 2013 Dec 4 15:22:41 & 120.0\\
I2 &I & 2013 Dec 4 15:25:10 &   2.0\\
I3 &I & 2013 Dec 4 15:25:51 & 120.0\\
I4 &I & 2013 Dec 4 15:28:29 & 120.0\\
V1 &V & 2013 Dec 4 15:39:57 & 120.0\\
V2 &V & 2013 Dec 4 15:42:33 &  2.0 \\
V3 &V & 2013 Dec 4 15:43:14 & 120.0\\
V4 &V & 2013 Dec 4 15:45:43 & 30.0 \\
V5 &V & 2013 Dec 4 15:46:43 & 10.0 \\
\hline
\end{tabular}
\caption{Observation Log}
\label{tab:obslog}
\end{table}

\begin{table}
\begin{tabular}{lc}
& JPL\#55\\
\hline
epoch& 2456653.5 (JD)\\
q    & 0.8118255522209098 \\
e    & 0.9984254232918264 \\
i    & 64.0409480407445 [deg]\\
w    & 67.16643021837888 [deg]\\
Node & 70.7111790783519 [deg]\\
Tp   & 2456649.2331241518 (JD)\\
\end{tabular}
\caption{Orbital Elements of C/2013 R1}
\label{tab:orbit}
\end{table}

\begin{table}
\begin{tabular}{|c|c|c|c|c|c|c|c|c|c|c|}
\hline
SDSS-Suprime& SDSS color& range &$c_0$&$c_1$&$c_2$ &$c_3$&$c_4$&$c_5$\\
\hline
$g-V$     & $g-r$ & -0.6$<g-r<$0.8 & 0.039 & 0.574 & -0.086 & 0.257 & 0.188 & -0.406\\
\hline
$i-I$     & $r-i$ & -0.4$<r-i<$0.5 & -0.018 & 0.255 & 0.037 & 0.092 & -0.196 & ...\\
\hline
\end{tabular}
\caption{The coefficients of color conversion}
\label{tab:colorcoeff}
\end{table}

\clearpage
\appendix

\section{Unsharp masking Procedure}

\begin{figure}
\includegraphics[scale=0.24,bb= 0 0 1910 928]{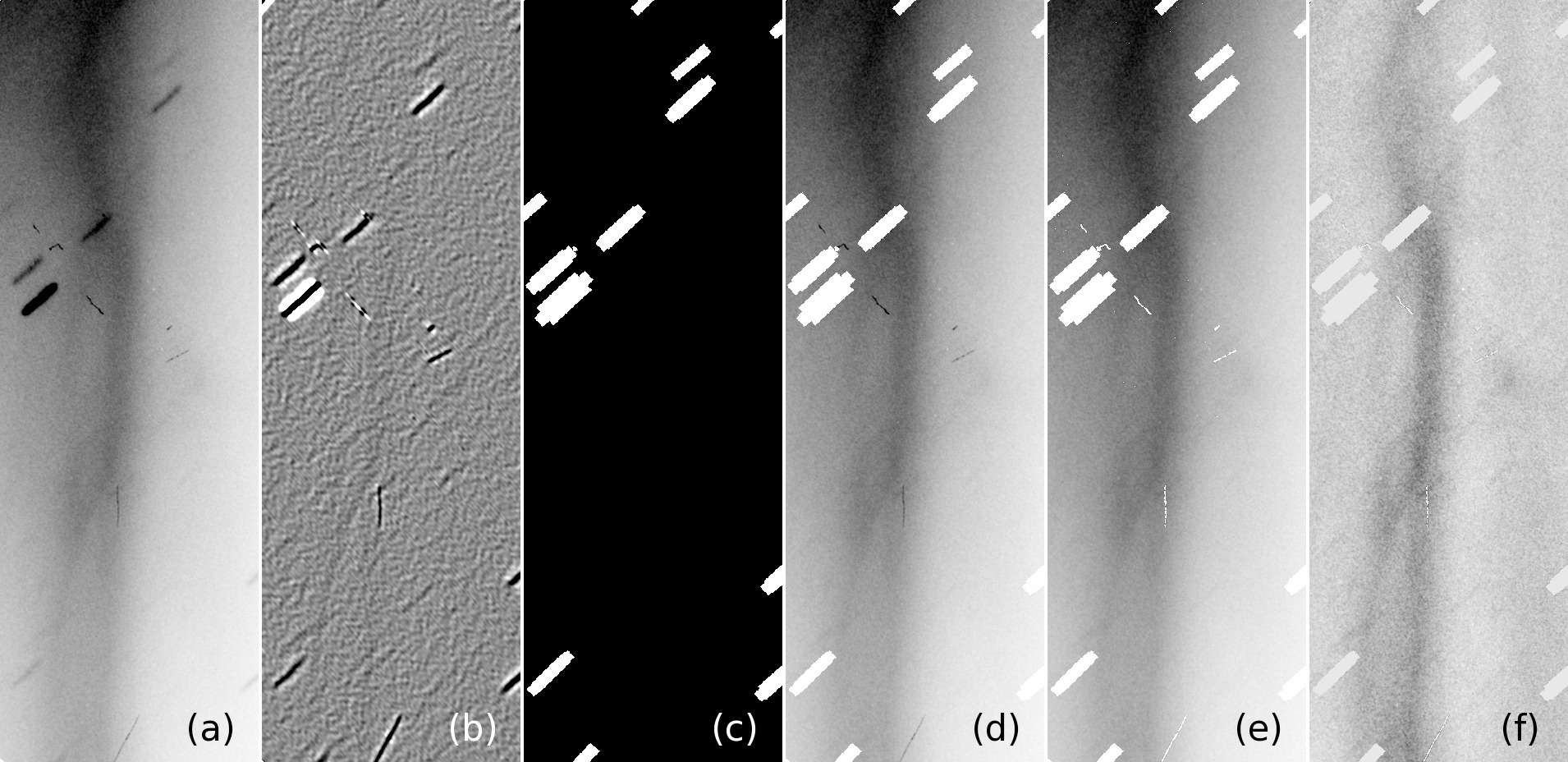}
\caption{An example of unsharp masking and star-masking process.
Cutouts of I3 around the knots are shown. 
The bright part is shown as black.
(a) Image after mosaicking, 
(b) Enhancement of star trails,
(c) Mask pattern made from (b). Mask is shown as white.
(d) ``AND'' mask of I1, I3 and I4. Cosmic rays and cometary structures are 
removed.
(e) Cosmic-rays mask added,
(f) Unsharp masked (e), using  $\sigma$=50 pixels Gaussian.
}
\label{fig:maskprocess}
\end{figure}

In this study, we first reduced the data following 
a general data reduction process 
using nekosoft \citep{Yagi2002}.
Figure \ref{fig:maskprocess}a shows a zoom-up of the area around
the the knots in the reduced image of I3.
The data consist of background objects 
(most of them are stars), 
and cosmic rays as well as 
the comet tails and their faint substructures.
We remove these background objects and cosmic rays in the following way.
The non-sidereal tracking generate a characteristic trail 
of background objects.
The direction and length of the trails are determined by 
the ephemeris that we adopted for the observations with the tracking.
To enhance the trails of the background objects
for an effective mask generation, 
we took the difference between two smoothed images, both of which were
smoothed with a two-dimensional Gaussian kernel with an elongation
perpendicular to the trails, but one with a smaller kernel with the
$\sigma$ of $1.5\times1.0$ pixels  and the other with a larger
kernel with $5.0\times2.0$ pixels.
This procedure generates a very recognizable pattern around 
objects that moved according to the adopted ephemeris.
Figure \ref{fig:maskprocess}b shows the pattern, 
and Figure \ref{fig:maskprocess}c shows 
a first mask generated from this image.
The background objects are fixed at celestial positions. 
We therefore shifted the first masks from subsequent exposures
and took an "AND" to make a final mask for all the exposures. 
The lengths of the trails depend on
exposure times, but the mask size would be comparable among the exposures
if the exposure time were the same.
We made one mask for I-band using (I1, I3, I4), 
and two masks for V-band using (V1, V3) and (V1, V3, V4)
-- the first V-band mask was applied for V1 and V3, and the second for V4,
since the shorter exposure image has a smaller trail of background objects.
An example of masked images is Figure \ref{fig:maskprocess}d.
In Figure \ref{fig:maskprocess}d, cosmic rays and bad pixels 
remain unmasked.
We then masked pixels whose value is larger than a threshold
(Figure \ref{fig:maskprocess}e).
We applied the unshaped masking technique to the images processed above. 
A result of the unshaped masking with $\sigma=50$ pixels (highpass50)
is shown in Figure \ref{fig:maskprocess}f, which is a part of Figure 2.

\clearpage

\section{Tests of detection and measurement with artificial images}

To test the reliability of detections of extended structures,
we employ tests with images of a fake object.
We adopt a circular Gaussian profile with a variety of widths,
i.e., $\sigma=$10, 20, 30, 40, 50, and 70 pixels, corresponding to the
FWHM of 4.7, 9.5, 14, 19, 24, and 33 arcsec, respectively. 
We set the peak of the Gaussian to be 70 counts, 
which is comparable to that of the two knots of interest 
discussed in Section \ref{sec:knots}.
Noise is also added in the artificial images 
using the double precision
SIMD-oriented Fast Mersenne Twister (dSFMT) software
\footnote{\url{http://www.math.sci.hiroshima-u.ac.jp/~m-mat/MT/SFMT/}}.

\subsubsection{Effect of noise}

We first examined the effect of relatively large noise, 
with S/N=1 at the peak of the Gaussian profile.
In this case, a 5$\times$5 binning improves the S/N to 5 at the peak. 
We ran SExtractor \citep{Bertin1996} for
object detection using the 5$\times$5 binned image.

Results are shown in Table \ref{tab:noiseobj}.
The second and third columns show measurements of 
input models before adding the noise (noise-free). 
The rest of the columns are for measurements of
detected objects using the images with the noise added (noise-added). 
We generated 1000 random realizations of noise pattern across the images. 
In some realizations, the Gaussian profile 
was incorrectly detected as blended objects
and was split into multiple objects. 
In those cases, we summed up the fluxes to calculate 
a total flux and measured the position of the
object using a flux-weighted mean. 
The FWHM size was not measured in case of 
the false multiple-object detection. 
Measured parameters in the two cases, i.e., 
detections of single and multiple objects are
separately shown in Table \ref{tab:noiseobj}. 
To estimate errors in FWHM and total flux, 
we used the median absolute deviation (MAD) 
and converted the MAD to rms as rms$\sim 1.48 \times$ MAD. 
This is known as a robust estimator of the rms and 
is valid for the normal distribution.

This test shows that the noise does not largely affect the position
measurement in case of single detection; only rms $\sim$ 0.23-0.24
pixels in 5$\times$5 binned image ($\sim$0.2 arcsec). 
In case of multiple
detections the rms of flux-weighted mean position error is up to 
0.80 pixels ($\sim$0.8 arcsec). 
The 0.8 pixel error corresponds to $\sim$ 300 km in this study, 
which is negligible (see Figure \ref{fig:speed}). 
The errors in FWHM and total flux measurements are not negligible.
In case of multiple-object detection, 
the FWHM is about 30-50\% larger than the measurements of 
noise-free models, which is significantly larger than
the rms of the measurements. 
The total flux is smaller by 0.06-0.15 mag in case of single detection, 
while it is larger by 0.11-0.33 mag in case of multiple detection. 
These differences are significant except for the $\sigma=$10 model.

The relatively large errors in FWHM and flux measurements may be due
to a possible overestimation of the background. 
We do not investigate the cause of the errors further 
as the quantitative error estimate given above 
is enough for this study.

\subsubsection{Effect of unsharp masking}

We also test errors due to the unsharp masking technique 
using the noise-added images. 
Measurements were made with the 5$\times$ 5 binned image
after application of the unsharp masking 
(highpass30, highpass50, and highpass70). 
We adopted the same 1000 noise realizations as in the previous section.

Results are shown in Table  \ref{tab:unsharp30}--\ref{tab:unsharp70}.
If the filter size $\sigma$ is larger than or comparable to a
$\sigma$ of Gaussian profile of the fake object, the unsharp-masking
technique does not cause large errors in measurements of position and
FWHM.
The rms of the positional shift is $<$ 0.5 pixels in case of single detection, 
and $<$ 0.7 pixels in case of multiple detections. 
The error, i.e., $<$0.7 pixels ($<$300 km), is negligible in this study. 
The errors in FWHM measurements are comparable 
before and after the unsharp masking, 
except for the case of highpass50 with $\sigma=$40. 
On the other hand,
if filter size $\sigma$ is smaller than 
that of Gaussian of the fake object, 
the deviation becomes larger and the fraction of no-detection increases.

As expected, the total fluxes measured after the unsharp masking are
significantly smaller than those before the masking. 
The difference is up to 1 mag in the case of single detections and up
to 2.4 mag in the case of multiple detections. 
The large error in flux, however, does not affect the analysis of this
paper, since our conclusion is based primarily on positional shifts
and corresponding velocities of the clumps. 
If we adequately select the filter size $\sigma$ with respect to the
sizes of the object, 
the unsharp masking technique works good for positional measurement.

\begin{table}
\begin{tabular}{c|cc|cc|ccc|cc}
\hline
          & 
\multicolumn{2}{c|}{noise-free}    &
\multicolumn{7}{c}{noise-added} \\
&
&&
&&
\multicolumn{3}{c|}{single detection} &
\multicolumn{2}{c}{multiple detection}\\
$\sigma$ & FWHM & total & multi. & no & 
shift rms & FWHM   & total &
shift rms & total \\
pixel   & pixel& count & det. & det.  & 
pixel     & pixel  & count &
pixel  & count \\
\hline
10 & 4.7 & 1800  &  4   & 0 & 
0.23 &7$\pm$1  & 1700  $\pm$200 &
0.26 & 2000$\pm$ 200 \\
20 & 9.4 & 7000  &  95  & 0 & 
0.23 &13$\pm$1 & 6500  $\pm$300 &
0.38 & 9500$\pm$ 1000 \\
30 & 14  & 15800 &  148 & 0 & 
0.23 &20$\pm$1 & 14300 $\pm$500 &
0.47 & 21300$\pm$ 1900 \\
40 & 19  & 28100 &  135 & 0 & 
0.24 &26$\pm$1 & 25100 $\pm$800 &
0.57 & 37600$\pm$ 3000 \\
50 & 24  & 43900 &  123 & 0 & 
0.24 &32$\pm$1 & 38800 $\pm$1000 & 
0.71 & 57000$\pm$ 5500 \\
70 & 33  & 85900 &  78  & 0 & 
0.24 &44$\pm$1 & 74700 $\pm$1700 &
0.80 & 105500$\pm$ 15200 \\
\hline
\end{tabular}
\caption{Measurement of artificial objects}
\label{tab:noiseobj}
\end{table}

\begin{table}
\begin{tabular}{c|cc|ccc|cc}
\hline
          & 
\multicolumn{7}{c}{highpass30} \\
&&&
\multicolumn{3}{c|}{single detection} &
\multicolumn{2}{c}{multiple detection}\\
$\sigma$ & multiple & no detection & 
shift rms & FWHM & total &
shift rms & total \\
pixel  &         &              & 
pixel & pixel & count &
pixel & count \\
\hline
10 &
 1 & 0 & 0.24 & 7$\pm$ 1 & 1200$\pm$ 100 &
 0.40 & 1200 \\
20 &
 61 & 0 & 0.30 & 12$\pm$ 1 & 2600$\pm$ 200 &
 0.44 & 3400$\pm$ 400 \\
30 &
153 &0  & 0.50 & 11$\pm$ 2 & 3100$\pm$ 300 &
0.52 & 4600$\pm$ 600
\\
40 &
 113 &181  & 1.01 & 8$\pm$ 2 & 2700$\pm$ 600 &
0.85 & 4200$\pm$ 700
\\
50 &
 14 &825  & 1.62 & 7$\pm$ 2 & 2100$\pm$ 600 &
1.16 & 3500$\pm$ 600
\\
70 &
 1 &996  & 3.94 & 8$\pm$ 3 & 700$\pm$ 300 &
2.90 & 1500
\\
\hline
\end{tabular}
\caption{Measurement of unsharp-masked artificial objects }
\label{tab:unsharp30}
\end{table}

\begin{table}
\begin{tabular}{c|cc|ccc|cc}
\hline
          & 
\multicolumn{7}{c}{highpass50} \\
&&&
\multicolumn{3}{c|}{single detection} &
\multicolumn{2}{c}{multiple detection}\\
$\sigma$ & multiple & no detection & 
shift rms & FWHM & total &
shift rms & total \\
pixel  &         &              & 
pixel & pixel & count &
pixel & count \\
\hline
10 &
 0 & 0& 0.23 & 7$\pm$ 1 & 1400$\pm$ 100 &
... & ... \\
20 &
 102 & 0& 0.24 & 13$\pm$ 1 & 4300$\pm$ 200 &
0.37 & 5700$\pm$ 500 \\
30 &
 264 & 0& 0.28 & 18$\pm$ 2 & 6900$\pm$ 400 &
 0.52 & 9700$\pm$ 1300 \\
40 &
 427 & 0& 0.36 & 18$\pm$ 3 & 8400$\pm$ 500 &
0.45 & 13300$\pm$ 1900 \\
50 &
 589 & 0& 0.48 & 15$\pm$ 4 & 9100$\pm$ 900 &
0.56 & 14900$\pm$ 2800
\\
70 &
 367 & 23& 1.98 & 8$\pm$ 2 & 4100$\pm$ 1500 &
 1.38 & 7700$\pm$ 2200
\\
\hline
\end{tabular}
\caption{Measurement of unsharp-masked artificial objects}
\label{tab:unsharp50}
\end{table}

\begin{table}
\begin{tabular}{c|cc|ccc|cc}
\hline
          & 
\multicolumn{7}{c}{highpass70} \\
&&&
\multicolumn{3}{c|}{single detection} &
\multicolumn{2}{c}{multiple detection}\\
$\sigma$ & multiple & no detection & 
shift rms & FWHM & total &
shift rms & total \\
pixel  &         &              & 
pixel & pixel & count &
pixel & count \\
\hline
10 &
 0 & 0&0.23 & 7$\pm$ 1 & 1600$\pm$ 100 &
... & ... \\
20 &
 110 & 0&0.24 & 13$\pm$ 1 & 5200$\pm$ 300 &
0.38 & 7200$\pm$ 600 \\
30 &
 242 & 0&0.25 & 19$\pm$ 2 & 9600$\pm$ 400 &
0.42 & 13400$\pm$ 1900 \\
40 &
 400 & 0&0.28 & 25$\pm$ 2 & 13500$\pm$ 600 &
 0.55 & 20300$\pm$ 3500 \\
50 &
 588 & 0&0.33 & 27$\pm$ 2 & 16300$\pm$ 100 &
 0.66\tablenotemark{1} & 26700$\pm$ 5100 \\
70 &
871 & 0&0.47 & 22$\pm$ 4 & 18400$\pm$ 1800 &
0.70 & 31900$\pm$ 6600
\\
\hline
\end{tabular}
\caption{Measurement of unsharp-masked artificial objects}
\label{tab:unsharp70}
\tablenotetext{1}{One outlier is removed.}
\end{table}

\end{document}